\documentclass[12pt]{article} 

\pdfoutput=1
\usepackage[margin=2cm]{geometry}

\usepackage[numbers]{natbib}
\usepackage[affil-it]{authblk}

\usepackage{amsmath}
\usepackage{amsthm}
\usepackage[english]{babel}
\usepackage{graphicx}
\usepackage{amsfonts} 
\usepackage{amssymb} 
\usepackage{hyperref} 
\usepackage{float}
\usepackage{times}
\usepackage{tensor}
\usepackage{mathtools}
\usepackage{color}

\numberwithin{equation}{section}

\newcommand{\deriv}[1]{\frac{d}{d #1}}
\newcommand{\DERIV}[2]{\frac{d #1}{d #2}}
\newcommand{\pd}[1]{\frac{\partial}{\partial #1}}
\newcommand{\PD}[2]{\frac{\partial #1}{\partial #2}}

\newcommand{\half}[0]{\frac{1}{2}}

\title{Quantum mechanics as the dynamical geometry of trajectories} 
\author{Philipp Roser}
\affil{Department of Physics and Astronomy,\\ Clemson University, Kinard Laboratory,\\ Clemson, SC 29631-0978, USA}
\date{\vspace{-0.5cm}}		

\begin{document}
 \maketitle
 \begin{abstract}
We illustrate how non-relativistic quantum mechanics may be recovered from a dynamical Weyl geometry on configuration space and an `ensemble' of trajectories (or `worlds'). The theory, which is free of a physical wavefunction, is presented starting from a classical `many-systems' action to which a curvature term is added. In this manner the equations of equilibrium de~Broglie-Bohm theory are recovered. However, na\"ively the set of solution precludes solutions with non-zero angular momentum (a version of a problem raised by Wallstrom). This is remedied by a slight extension of the action, leaving the equations of motion unchanged. 
\end{abstract}

\section{Introduction}\label{introduction}
The first of two recent but entirely independent developments in the foundations of quantum theory is a number of `trajectory-only' formulations of quantum theory \citep{Sebens2014, HallDeckertWiseman2014, Tipler2010, Bostrom2014, Poirier2010, SchiffPoirier2011, Schmelzer2011a}, though also see \citep{Holland2005}, intended to recover quantum mechanics without reference to a physical wavefunction. Two related approaches which however leads to an experimentally distinguishable theory are the real-ensemble formulations of \citep{Smolin2011,Smolin2015}. The other recent development of importance to us is the insight that the non-classical nature of quantum dynamics may be understood as a geometrical phenomenon, as has been shown for non-relativistic particle mechanics \citep{Santamato1984, Carroll2007, NovelloSalimFalciano2009}, relativistic particle mechanics \citep{Santamato1984b,Santamato1985} and for the particle Dirac equation \citep{SantamatoMartini2011,SantamatoMartini2012,SantamatoMartini2014}. In this paper we demonstrate that these two ideas form a natural union, and that this union may indeed help to overcome a technical problem of the trajectory-only approach that has remained largely, though not entirely, unaddressed.

We will show how quantum theory may be understood as a natural extension of classical mechanics by allowing the configuration-space geometry to be dynamical. Here we focus on the non-relativistic theory. In previous approaches to such trajectory-only formulations quantum mechanics was recovered by postulating a `force' term that was Newtonian in the sense that it appeared on the right-hand side of `$m\ddot{x} = $' (but was not Newtonian in any other sense of the word). In its simplest formulation \citep{Sebens2014} this term was chosen to match the quantum potential of Bohm \citep{Bohm1952a, Holland1993} with the substitution $|\psi|\rightarrow\sqrt\rho$. That is, the Born probability amplitude $|\psi|$ was replaced by the square root of the system density $\rho$ in configuration space. This eliminates any reference to a wave function $\psi$, which is not part of such a trajectory-only theory. However, it does require that $\rho$ is a physically real quantity. This means it requires the reality of \emph{all} dynamically allowed configuration-space trajectories, making it a theory of many worlds.\footnote{Some authors in their presentation have therefore likened aspects of the many-trajectory approach to Everettian `many-world' theory. We find this more confusing than helpful since the two have very little in common with the exception of being described as theories of `many worlds', a phrase with rather distinct meanings in the two theories.}

In this paper we take a different approach and do not postulate such a force. Instead, we begin by reconsidering aspects of classical mechanics, constructing an action that leads to a simultaneous determination of all dynamically possible paths. Mathematically this is almost trivial given the action for a single trajectory. However, it prepares the next step, the introduction of a geometrical term in the action. This term is nothing but the curvature of the configuration space. So the action to be associated with quantum mechanics will be the sum of two terms: The classical matter action plus a curvature. Formally it is similar to the Einstein-Hilbert action, although this analogy should not be taken too literally since the latter concerns the curvature of ($3+1$)-spacetime, not configuration space.

The action we obtain is effectively that of Santamato \citep{Santamato1984}, but unlike Santamato's it does not rely on the minimisation of expectation values. Nonetheless a substantial part of our derivation is taken from \citep{Santamato1984}. Our action differs from that of \citep{NovelloSalimFalciano2009} in its functional form, although it turns out to be numerically equal on shell.

The trajectories we find turn out to be exactly those of de~Broglie-Bohm theory \citep{Bohm1952a, Holland1993, deBroglie1928, Bohm1952b} for an ensemble in equilibrium ($\rho=|\psi|^2$).\footnote{It is easily provable that if this condition holds at one time $t_0$, then it will hold at all times $t>t_0$. That the ensemble density matches $|\psi|^2$ at some initial time $t_0$ is considered a postulate of de~Broglie-Bohm by some authors, possibly justified by some kind of typicality argument \citep{DuerrGoldsteinZanghi1992}. However, this is unnecessary since it can be shown that in reasonable circumstances an ensemble out of equilibrium ($\rho\neq|\psi|^2$) rapidly approaches equilibrium \citep{Valentini1991a, ValentiniWestman2004}.}
Equilibrium de~Broglie-Bohm theory exactly recovers all experimental predictions of quantum mechanics, and so the theory presented here does so too. There is, however, a subtlety , which we will address.

The dynamical geometry of the configuration space is a so-called \emph{Weyl geometry} \citep{Weyl1918}, of which we will give a brief review in section \ref{Weyl}. We will find that the geometry has singularities, namely at points which in conventional quantum mechanics (or de~Broglie-Bohm theory) correspond to nodes in the wavefunction ($\psi=0$). This will lead us to refine our approach: We will move from trajectories \emph{in} the configuration-space manifold to a manifold \emph{of} trajectories. This refinement is crucial to what is to follow, as well as strongly preferable for philosophical reasons. However, for reasons of presentational clarity we will only make this change in our perspective once the basic tenets of our theory are established. 

As the final step in our recovery of the phenomenology of quantum mechanics we address the aforementioned `subtlety', namely the quantised nature of angular momentum, manifest in the fact that loop integrals of the form $\oint \vec{p}\cdot\vec{dl}$ may be non-zero, yet only take values $2\pi\hbar\cdot m$, $m\in\mathbb{Z}$. In standard de~Broglie-Bohm theory this result follows from the multi-valuedness of the phase of the wavefunction and the fact that the phase is undefined at nodes. In a na\"ive trajectory-only theory it is entirely unclear why such a result should hold, or how the integral could be non-zero at all. This was first pointed out by Wallstrom \citep{Wallstrom1994} in connection with hydrodynamical approaches and has remained largely unaddressed.\footnote{Strictly speaking only the first of these questions is the one asked by Wallstrom in \citep{Wallstrom1994}. However, I will use the phrase `Wallstrom's objection' more broadly and include also the second where applicable.}

Coming from the perspective of de~Broglie-Bohm, the idea of `many de~Broglie-Bohm world' theory may be considered a precursor to the trajectory-only theory. Here an infinite ensemble of worlds evolves according to the laws of standard de~Broglie-Bohm theory, guided by an ontological wave function. The notion appeared inadvertently as a result of the `many-worlds-in-denial' objection to de~Broglie-Bohm theory \citep{Deutsch1996,BrownWallace2005} and was discussed and criticised by Valentini \citep{Valentini2010}, whose position was subsequently challenged by Brown \citep{BrownReplyToValentini2010}. Since there is a physical wave function in this theory, Wallstrom's problem does not arise. The concept is also discussed by Sebens \citep{Sebens2014}, who calls the theory `prodigal QM', although he considers it with only a finite number of trajectories.

Let us now turn to many-trajectory theory proper. It is instructive to outline the approaches of other authors briefly and review how they have addressed the Wallstrom objection, before summarising the solution presented in this paper. This short survey is not intended to be comprehensive or detailed.
 
The idea of formulating quantum mechanics in terms of the flow of an ensemble of configuration-space particles is sometimes traced back to ideas of Madelung \citep{Madelung1926}. However, this is a rather anachronistic reading of Madelung and historically inaccurate \citep{BacciagaluppiValentini2009}: Madelung imagined the electron as a charged fluid in three-space and he did not make the leap to \emph{configuration space} as the arena of the dynamics, nor did he consider individual trajectories within the fluid to have any ontological significance. His approach is therefore more akin to that of Schr\"odinger than having any resemblance to the trajectory-based approached with which we are concerned here. 

Sebens \citep{Sebens2014} arrives at the trajectory-only theory by starting from the point of view of de~Broglie-Bohm, introducing an ensemble of `worlds' distributed according to $|\psi|^2$ and then eliminates $\psi$ in favour of the ensemble density. However, he suggests that this is only a continuum limit of a more fundamental ontology with only a finite (though large) number of worlds. As such, quantum mechanics is considered to be only approximately correct, though sufficiently accurate for a difference to be experimentally undetected at this point. Sebens addresses Wallstrom's objection by introducing an appropriate quantisation condition, although stating that this is preliminary and that for now the quantised nature of these loop integrals is an empirically discovered feature of our universe. 

The approach of Hall et al.\ \citep{HallDeckertWiseman2014} is similar to that of Sebens, except that they discuss in more detail the nature of the inter-world interaction in the case of a finite number of worlds and indeed provide a concrete example for a one-dimensional model. They also apply the theory to a series of quantum phenomena and use it as a tool to numerically approximate ground state wavefunctions. They do not address Wallstrom's objection however.

Tipler \citep{Tipler2010} (but also see his earlier work \citep{Tipler2006}) begins with classical mechanics with an infinity of trajectories and then introduces the `smoothing potential' $U$ (equal to the quantum potential once the function $R$ appearing in the definition of $U$ has been made) as the unique correction to the Hamilton-Jacobi equation that avoids the crossing of trajectories and thereby ensures the smoothness of Hamilton's principal function $S$ and, in his words, prevents `god from playing dice', referring to the ambiguity in evolution where trajectories would cross in the absence of the smoothing potential. He likens this addition to Maxwell's introduction of the displacement current in Amp\`ere's law. The constant $\hbar$ must exist based on dimensional grounds. He also discusses the recovery of Born probabilities and the uncertainty principle in terms of his theory. However, despite employing an example involving angular momentum he does not address Wallstrom's issue. Indeed, $S$ is a real valued function by construction ---the classical Hamilton principal function--- and therefore any non-zero value of angular momentum remains to be explained.

A critique of Tipler's proposal can be found in Bostr\"om \citep{Bostrom2014}. Bostr\"om's approach shares features with that of Sebens, Hall et al.\ and Tipler, but he is very careful about discussing the philosophical and logical foundations required for the construction of his theory. Somewhat peculiarly though the existence of a wave function is one of the postulates of his theory, although he makes clear that `existence' here is meant in a mathematical sense (that is, such a function is definable) and not in some ontological sense. This does help his theory to overcome Wallstrom's objection, although the theory may be seen as somewhat unsatisfactory given the reliance on a wave function, ontological or not. Unlike Sebens and Hall et al., Bostr\"om argues in favour of a continuous infinity of worlds rather than a large but finite number and he briefly discusses the measure-theoretic foundations for such a picture, in particular in response to criticism by Vaidman \citep{Vaidman2014}.

The approach of Schiff and Poirier \citep{Poirier2010, SchiffPoirier2011} is different in that their formulation does not rely on the introduction of a quantum potential in the usual form. Instead, they consider higher-order time derivatives of the variables to appear in the expressions for the Lagrangian and energy of their theory. Certain constraints on their form then allow the identification of a term that effectively constitutes (at least numerically) the quantum potential as the simplest option (although at least in the time-dependent case reference to a density of trajectories in configuration space cannot be avoided). The authors show how the function $S$ can be recovered as the action integral along a trajectory, but why this function should indeed be phase-like in its multivaluedness remains unaddressed.

Schmelzer's \citep{Schmelzer2011a, Schmelzer2011b} construction is somewhat similar to Tipler, Sebens and Hall et al., although he discusses a number of foundational points not addressed by them and, crucially, provides a detailed discussion of Wallstrom's objection together with a proposal to overcome it. He shows that a regularity postulate for the Laplacian of the density $\Delta\rho$ suffices to restrict the value of the loop integrals $\oint \vec{p}\cdot\vec{dl}$ to take values equal to $\pm\hbar$. The postulate is equivalent to demanding that the nodes of the wave function are simple zeros. Since other integer values are however possible in standard quantum theory such situations must correspond to the presence of multiple zeros within the loop. For this to be viable one must consider quantum mechanics to be merely an approximation to some underlying `subquantum' theory, whose exact nature is unknown, but which may be empirically distinguishable, at least in principle. Therefore, while not proposing an experimentally distinct theory, Schmelzer's reasoning may nonetheless be experimentally assessible, although hard to ever rule out since, if standard quantum mechanics holds up, one is always able to change the regime (scale) on which differences are expected to become detectable. Despite this departure from quantum mechanics, Schmelzer does resolve the Wallstrom objection in his framework. The regularity postulate itself is justified by another principle concerning the nature of the elusive subquantum theory, which he calls the `principle of minimal distortion', together with a recognition of the role of $\Delta\rho$ in the energy-balance equation associated with the trajectories, although he acknowledges that this specification is `speculative in character' \citep{Schmelzer2011b} because it concerns the unknown subquantum theory.

Smolin \citep{Smolin2011,Smolin2015} has made two proposals for \emph{real}-ensemble formulations, meaning formulations of quantum mechanics in terms of ensembles of identical systems in a single (our) world. As such they do not belong to the family of many-trajectory theories discussed so far. However, Smolin's theories address Wallstrom`s objection and are therefore of interest to our purposes, although we cannot, of course, do his approaches justice here. His first formulation \citep{Smolin2011} is based on a stochastic dynamics governing how systems copy their beables onto those of similar system (in addition to a further deterministic dynamical law concerning a `phase' associated with each system). Wallstrom's objection is avoided in virtue of the form of the action, in which the phase variables appear as phases, not their real-valued gradient. There are other issues with this approach however, notably what the author calls `phase alignment', for which a mechanism is suggested but which, in the author's own words, ``is ad hoc and can probably be improved on''. His second formulation \citep{Smolin2015} does not rely on a stochastic mechanism but the `principle of maximal variety', which effectively drives systems towards greater distinctiveness (a term which is given a quantitative definition) with regards to their relation to the rest of the universe. Mathematically this is achieved via a potential-energy term that is a function of the ensemble properties, in particular the aforementioned `distinctiveness'. In the appropriate limit quantum mechanics can be recovered, although one can expect experimental deviations from quantum mechanics for appropriately rare systems, such as macroscopic objects. Motivation for this idea is found in Leibniz' Principle of the Identity of Indiscernibles. Smolin addresses Wallstrom's issue by introducing unimodular complex beables to take the role of momenta, rather than real-valued phases. This is a relatively straightforward solution, although it may be criticised on the basis that it was introduced specifically to solve this problem and without independent motivation.

Our solution to the Wallstrom's objection consists of a slight generalisation of our action, bringing it into the most general form that still retains the same equations of motion. While this modification is, of course, prompted by the necessity to address Wallstrom's objection, we emphasise that the new term is by no means ad hoc. Rather, it is the most general total derivative that can be consistently added to the Lagrangian and the addition of the total derivative in the first place is connected to the open-endpoint variation employed in the construction (see sec.\ \ref{classical}). 

We end with some remarks on field theory. After all, the physical world is manifestly not governed by non-relativistic quantum mechanics and the real test of the viability of our approach is a reconstruction of field-theoretical results.


\section{Aspects of classical mechanics}\label{classical}

In this section we discuss two somewhat unconventional aspects of classical mechanics that will be relevant when we introduce our action for quantum trajectories below. The first of these is variation of an action without fixing the end point, apparently leading to the condition that the momenta $p_i$ vanish. However, if the Lagrangian is specified only up to a total derivative of some unknown function $S$ (this does not change the equations of motion), then the condition obtained is $p_i=\partial_iS$. This use of open-endpoint variation is due to Santamato \citep{Santamato1984}. Tipler \citep{Tipler2010} uses open-endpoint variation in a similar though not identical manner in his derivation of the Hamilton-Jacobi equation. The second point of discussion is the simultaneous variation of multiple trajectories in configuration space by summing (or integrating) the actions of each individual system. At first this remark may appear almost trivial, yet it will present a natural starting point for our derivation of quantum trajectories later on.

\subsection{Open-endpoint variation and total derivatives}
Let us assume that in our classical theory the trajectories $q(t)$ traced out in a configuration space $\mathcal{C}$ may be obtained by extremisation of an action,
\begin{equation} I_1 = \int dt\; L(q(t),\dot{q}(t),t) \end{equation}
where $q=\{q^1, \dots, q^n\}$ are the configuration variables and $L$ is the Lagrangian. A dot denotes differentiation with respect to the time parameter $t$. In the usual approach one fixes the two end points of the trajectory, that is, the configuration of the system at two times $t_0$ and $t_1$: $q_0=q(t_0)$ and $q_1=q(t_1)$, where $t_1>t_0$. Since the value of $I_1$ (and $L$ and the intermediate points $q(t)$) depends on the choice of end points, we should strictly write
\begin{equation} I_1(q_0,t_0;q_1,t_1) = \int_{t_0}^{t_1} dt\; L\big(q(q_0;q_1;t),\dot{q}(q_0;q_1;t),t\big).\end{equation}
This is usually not done since it is cumbersome notation and the choice of end points $q_0$ and $q_1$ does not affect the equations of motion. Only the functional form of the Lagrangian is of significance. Rarely is the numerical value of the action integral of interest.

The variation $\delta q$ at $t_0$ and $t_1$ is assumed to vanish, and so there is no boundary term when performing the partial integrations that lead to the equations of motion. But consider now the case where we only set $\delta q=0$ at $t_0$, but leave it open at $t_1$.\footnote{We could have chosen to leave both open, or only the initial variation. The reasoning that follows would be similar.} We therefore write
\begin{equation} I_1(q_0,t_0;t_1) = \int_{t_0}^{t_1} dt\; L\big(q(q_0;t),\dot{q}(q_0;t),t\big).\end{equation}
In that case variation yields a non-zero boundary term, proportional to the variation at $t_1$:
\begin{align}\notag \delta I_1(q_0,t_0;t_1) =& \PD{L}{\dot{q}^i}\big(q(t_1,q_0), \dot{q}(t_1,q_0),t\big) \cdot\delta q^i(t_1,q_0)  \\
				 &\;-\int\limits_{t_0}^{t_1}dt\, \left[\pd{t}\PD{L}{\dot{q}^i}\big(q(t,q_0),\dot{q}(t,q_0),t)\big) 
				    - \PD{L}{q^i}\big(q(t,q_0),\dot{q}(t,q_0),t\big) \right] \cdot\delta q^i(t,q_0).
\end{align}
The variation $\delta q(t_1,q_0)$ at the end point is independent of the variation $\delta q(t,q_0)$ at intermediate times. Hence an extremisation of $I_1$ implies that the term multiplying $\delta q(t_1,q_0)$ must vanish, as does the expression inside the square brackets. The latter gives the usual Euler-Lagrange equation, while the former implies that 
\begin{equation} 
    p_i\big(q(t_1,q_0), \dot{q}(t_1,q_0),t\big)\equiv\PD{L}{\dot{q}^i}\big(q(t_1,q_0), \dot{q}(t_1,q_0),t\big) = 0.\label{eq:RawBoundaryEq.}
\end{equation}
The momenta at the end point vanish. In fact, the appearance of such an end-point constraint is a general feature of variational problems with open boundary conditions \citep[chap.~2]{Lanczos1949}. 

In the context of Lagrangian mechanics this condition appears rather undesirable, as it seems to suggest that `ignorance' of the final position of our system implies that its momenta must vanish (and if our action principle is be extremised for \emph{any} choice of $t_0$ and $t_1$ the implication is that the momenta would have to vanish at all times). Nonetheless open-endpoint variation has found to be useful in the elimination of unphysical (gauge) degrees of freedom via a procedure called `best-matching' \citep[Sec. 4]{Barbour2003SIGPD}.

But a slight generalisation of our Lagrangian alleviates any concern we might have regarding the boundary equation \ref{eq:RawBoundaryEq.}. It is well known that the equations of motions are unchanged if a total time derivative of an arbitrary function $S(q,t)$ of the system variables and time is added to the Lagrangian:
\begin{equation} \bar{I}_1(q_0,t_0;t_1) = \int_{t_0}^{t_1} dt\; \left[L\big(q(q_0;t),\dot{q}(q_0;t),t\big)-\DERIV{S}{t}\big(q(q_0,t),t\big)\right].\end{equation}
The relative minus sign is for later convenience. Note that $\DERIV{S}{t}=\PD{S}{t}+\dot{q}^i\partial_iS$. Performing the variation as before, the boundary equation is now
\begin{equation}
  p_i\big(q(t_1,q_0), \dot{q}(t_1,q_0),t\big)-\partial_iS\big(q(t_1,q_0),t\big) = 0, \label{eq:BoundaryEq.}
\end{equation}
where $p_i$ is defined as above (that is, in terms of the original Lagrangian without the total time derivative). Since this is to hold for any interval $[t_0,t_1]$, we have in general that $p_i=\partial_iS$. This equation can be seen as determining $S$ up to an arbitrary additive function of time.\footnote{If we really wish to use the most general Lagrangian that has the same equations of motion, then another subtlety arises regarding the function $S$. We will not go into it here though but reserve it for section \ref{Wallstrom}, where it will be of relevance.} It only restricts the momenta $p_i$ in that these must be the gradient of \emph{some} function.

Of course, $S$ turns out to have all the properties of a generator of those canonical transformations that represent the time evolution of the system. In a more usual approach we might have introduced it as such. However, here we see that such a function can also be introduced in the Lagrangian picture as an arbitrary addition to the Lagrangian that leaves the equations of motion unchanged. For example, using $\dot{q}^ip_i=\dot{q}^i\partial_iS=\DERIV{S}{t}-\PD{S}{t}$, together with $H(q,\nabla S,t)\equiv\dot{q}^i\partial_iS-L(q,\dot{q},t)$ allows one to obtain the usual Hamilton-Jacobi equation for our function $S$.

For readers familiar with the use of open-endpoint variations in the context of 'best-matching' it is worth noting that Barbour's and his collaborators' use \citep{Barbour2003SIGPD} and ours is perfectly consistent, provided we demand that the function $S$ depend only on physical and not on non-physical gauge degrees of freedom. Then the momenta associated with the non-physical variables vanish, while those associated with physical ones satisfy $p_i=\partial_iS$. In fact, we note here as an aside that we believe this to be a promising line of investigation for the construction of a fully relational quantum theory from first principles.


\subsection{Action principles for multiple simultaneous trajectories}
The extremisation of $I_1$ (or $\bar{I}_1$) determines dynamically allowed configuration-space trajectories of a single system, starting at $q_0$ at time $t_0$. In the absence of a fixed end point, a second boundary condition (such as the initial velocity $\dot{q}(t_0)$) must be given in order to pick out a single trajectory uniquely.

Suppose now that instead of one such system we have $N$ systems of the same type, located at positions $q_{0(1)},q_{0(2)},\dots$ at start time $t_0$, with $q_{0(J)}=\{q_{0(J)}^1,\dots,q_{0(J)}^n\}$, $J=1,\dots,N$. The usual thing to do would be to extremise $\bar{I}_1(q_{0(J)},t_0;t_1)$ for each $q_{0(J)}$. Each time we would get the same equations of motion and only the initial conditions differ. That is, we have a different action for each system (taking different numerical values), although all the actions are identical in their functional form. This latter fact is the reason that in practice we would only derive the equations of motion once and then merely introduce the boundary conditions when solving them.

However, it is also possible to extremise a single action in order to obtain all $N$ trajectories at once (provided sufficiently many boundary conditions are specified). To do so we use the action
\begin{equation} I_N(q_{0(1)},\dots,q_{0(N)},t_0;t_1)=\frac{1}{N}\sum\limits_{J=1}^N I_1(q_{0(J)}).\end{equation}
The factor $1/N$ is not necessary for finite $N$ but is introduced here for consistency with what is to follow. The variations $\delta q(t,q_{0(J)})$ for each $J$ are independent of one another and so this action yields $N$ sets of formally identical Euler-Lagrange equations. In addition we have $N$ sets of boundary equations \ref{eq:BoundaryEq.}, one for each system.


We can generalise this to an infinite number of systems. For a countable infinity we take the limit,
\begin{equation} I_\infty = \lim_{N\rightarrow\infty} I_N(q_{0(J)};J\in\mathbb{N}).\end{equation}
Here the factor $1/N$ is important since otherwise the numerical value of the action would in general be infinite and thus not extremisable.\footnote{Alternatively we could have written down an infinite sum explicitly, but then we must use coefficients $\alpha_J$, which converge to zero sufficiently quickly as $J\rightarrow\infty$, so that the sum $\sum_{J=1}^\infty \alpha_J I_1(q_{0(J)})$ is finite. However, this would give different `weights' to different systems.} Here we will not consider the countable case any further. 

For a continuous (uncountable) infinity of system we must define a measure $\mu(q_0)$ over the configuration space, representing the density of systems at $q_0$ at time $t_0$. Since $\mu$ is defined over the whole configuration space (with $\mu=0$ in places where there are no systems), the starting positions $q_0$ take on the role of \emph{coordinates}. Let us therefore make the purely notational replacement $q_0\rightarrow x$. Instead of the discrete label $J=1,2,\dots$, systems are now labelled by the \emph{value} of their initial position $x$ (or $q_0$). The total action for this infinite ensemble of systems is now an integral over all starting positions $x$ with measure $\mu(x)$, which we assume to be normalisable. The action is therefore
\begin{align} I_{Ens} 
 &= \int_\mathcal{C} d^nx\,\mu(x) I_1(x,t_0;t_1) \notag\\
 &= \int_\mathcal{C} d^nx\,\mu(x) \int_{t_0}^{t_1} dt\, \left[L\big(q(t,x),\dot{q}(t,x), t\big)-\DERIV{S}{t}\big(q(x,t),t\big)\right]
\end{align}

Let us reiterate: $x$ labels the initial position at time $t_0$ and therefore functions as a coordinate for an ensemble whose distribution in $\mathcal{C}$ forms a continuum (or at least may be approximated by one). The dynamical variable itself is $q(t,x)$, that is, the variables are labelled by their initial position $q(t_0,x)=x$. The integral of $x$ over the configuration space $\mathcal{C}$ is not an integral of the dynamical variable but of the initial positions. (It would be non-sensical to integrate over all $q$ and at the same time attempt to arrive at dynamical laws for $q$.) 

Note that $\mu(x)$ is a function over the configuration space describing the ensemble density at time $t_0$, not over the variables $q$. This measure turns the integral over configuration space into an integral over all systems in the ensemble. If the ensemble is allowed to evolve for some time, and then the systems are `relabelled' at some later time $t_0^\prime$ by their instantaneous positions $q(t_0^\prime,y)=y$, the new density function $\mu^\prime(y)$ will of course not be identical with the original $\mu(x)$.

One may furthermore show using standard results of classical mechanics that the measure $\mu$ obeys a continuity equation,
\begin{equation} \PD{\mu}{t}+\partial_i\left(\dot{q}^i\mu\right) =0. \label{eq:ClassicalCE1}\end{equation}
This follows from the fact that the momentum $p_i$ is given by the spatial gradient of a function $S$, which ensures that trajectories can never cross (or begin or end) and furthermore are continuous on the configuration space.

We are now in a position to make the step towards the quantum theory.

\section{Review of Weyl geometry}\label{Weyl}
In this short section we present the main ideas and relevant results of a Weyl geometry. 
The primary source (and in our opinion still the best one despite its age) is found in the mathematical sections of Weyl`s own work \citep{Weyl1918}.\footnote{The sections on physical interpretation in Weyl's paper are also of extraordinary interest, although it seems that the (rather beautiful) idea to identify the Weyl one-form with the electromagnetic 4-potential ultimately fails. See the Einstein-Weyl correspondence following Weyl's other 1918 paper \citep{LorentzEtAl1923}.} Here we will remain brief and only quote the results relevant to us. 

A Weyl geometry may be seen as a generalisation of a Riemannian geometry in the sense that it is concerned with manifolds equipped with a metric $g_{ij}$. They key difference to a Riemannian geometry is that `length' is not comparable across finite distances. In Weyl's words, it is a `pure infinitesimal geometry'. Consider a vector $A^k$ at some point $P$ in the manifold, and suppose the vector is parallel-transported along some curve $C$ to a point $P^\prime$, where the transported vector is denoted by $A^{\prime k}$. In the case of a Riemannian geometry the direction of the vector is dependent on the connecting curve $C$. However, its length is not. The length $|A^\prime|=\sqrt{g_{ij}A^{\prime i}A^{\prime j}}$ depends only on the local metric $g_{ab}(P^\prime)$. In this sense length is path-independent and can be compared across finite distances in a Riemannian manifold. 

This is not the case if the geometry is a Weyl geometry. Here the length $|A^\prime|$ of the transported vector $A^k$ is also path dependent. For reasons described in \citep{Weyl1918} it is arguably natural to demand that for an infinitesimal transformation $P\rightarrow P+dP$, or in coordinates $x^i+dx^i$, the change in length is linear. That is, for a general length $\ell$, we find
\begin{equation} \ell+d\ell = \ell+ \ell\phi_kdx^k, \end{equation}
where $\phi_k$ is an arbitrary one-form and the full metric properties of the manifold are given by the pair $(g_{ij},\phi_k)$. 

In general, an infinitesimal transport results in a change in the vector given by $\delta A^k=\Gamma^k_{ij}dx^iA^k$, where the coefficients $\Gamma^k_{ij}$ are a priori arbitrary. However, this definition of $\Gamma^k_{ij}$, together with $\delta |A|^2 = \delta (g_{ij}A^iA^j)$, implies that
\begin{equation} \Gamma^k_{ij} = \half g^{kl}\left(\partial_ig_{jl}+\partial_jg_{il}-\partial_lg_{ij}\right)
				    +\half g^{kl}\left(g_{li}\phi_j + g_{lj}\phi_i - g_{ij}\phi_l\right),
\end{equation}
If $\phi_k\equiv0$, this expression reduces to the usual Christoffel symbol and the geometry is Riemannian. That is, the vectors undergo no change in length as they are transported along a curve.

Curvature is defined exactly as in the Riemannian case, but using the connection $\Gamma^k_{ij}$ obtained here. After some algebra we see that the scalar curvature (the one pertinent to our action) is given by
\begin{equation}
 R = R_{Riem} + (n-1)(n-2)\phi_k\phi^k-\frac{2}{\sqrt{g}}(n-1)\partial_k(\sqrt{g}\phi^k).
\end{equation}
Also note that the covariant derivative (defined in the usual way) of the metric does not vanish:
\begin{equation}
 \nabla_kg_{ij} = g_{ij}\phi_k.
\end{equation}
This implies that if one chooses coordinates such that locally the metric is $g_{ab}=diag(1,1,\dots)$ it is \emph{not} the case that the connection coefficients $\Gamma^a_{bc}$ vanish.\footnote{Contrast this to General Relativity, where $\nabla g_{\mu\nu}=0$, so that Lorentz frames (where $g_{\mu\nu}=diag(-1,1,\dots)$ locally) are inertial frames (where $\Gamma^\alpha_{\mu\nu}=0$) and vice versa (see \citep{MisnerThorneWheeler1973})}

\section{Quantum trajectories from an action principle}\label{quantum}
In this section we allow the geometry of the configuration space to become dynamical. To this end, we treat the Weyl one-form $\phi_k$ as a dynamical variable and add to the classical ensemble action $I_{Ens}$ a curvature term with an appropriately chosen constant of a form closely resembling conformal coupling. Extremising this action we obtain the equation of motion for $q$ (now with an extra non-classical term stemming from the curvature term), the continuity equation (from variation of the function $S$) and the relationship between the geometry and the measure $\mu$. We show how these equations reproduce quantum behaviour, up to a subtlety that we will address in section \ref{Wallstrom}.

\subsection{Preliminary remarks}
In our discussion in section \ref{classical} the density $\mu(x)$ did not play any dynamical role since in the extremisation of the ensemble action $I_{Ens}$ the total minimum was found as the weighted sum of the minima of each trajectory. That is, the minimisation of one trajectory was independent from the minimisation of any other. We will now introduce a term that effectively constitutes an interaction between trajectories. More precisely, we add the curvature of the configuration space to our Lagrangian (with a suitably chosen coupling constant that essentially turns out to correspond to $\hbar^2$ up to a numerical factor resembling a conformal coupling). This curvature is that of a Weyl geometry (see section \ref{Weyl} and \citep{Weyl1918}) and its Weyl one-form $\phi_k$ is considered to be a dynamical variable. The role of the `Riemannian' part of the curvature (due to the metric $g_{ij}$) is yet to be investigated. Here it is of secondary importance and $g_{ij}$ is not considered dynamical for our current purposes. 

Much of the mathematical structure appearing in this construction is again due to Santamato \citep{Santamato1984} and our notation is close to (though not identical with) his. However, Santamato's approach is founded on the stochastic approach of Nelson and Madelung, not a many-trajectories theory. Santamato's quantum action also involves the minimisation of an expectation value, which one might find ontologically questionable.\footnote{How can we make sense of the concept prior to having a theory of measurement and without the notion of quantum states, for example?} In our case, the minimisation principle remains unchanged from the classical minimisation of $I_{Ens}$. Merely an interaction term is added in the form of the curvature term. This curvature term constitutes an interaction between `neighbouring' trajectories since it contains configuration-spatial derivatives.

Note that our (and equivalently Santamato's) action is numerically identical to the action used in \citep{NovelloSalimFalciano2009} after use of $\DERIV{S}{t}=\PD{S}{t}+\dot{q}^i\PD{S}{q^i}$, $p_i=\PD{S}{q^i}$ and $H=\dot{q}^ip_i-L$. However, we consider our action to be a more natural starting point as a natural extension of the classical action. The approach of \citep{NovelloSalimFalciano2009} also differs in three other regards: Firstly, the authors are committed to consider a geometry on space rather than configuration space and their approach therefore only applies to a single particle (although without this commitment their formalism easily generalises). Secondly, the variation of their curvature term is done in the style of Palatini, that is, by considering the connection $\Gamma^a_{bc}$ as independent variables. Here on the other hand we will vary the curvature $R$ as a function of the Riemannian metric $g_{ij}$ and Weyl one-form $\phi_k$, although this latter difference is primarily aesthetic. The final difference is that they presuppose the Weyl geometry to be integrable, whereas in our formalism (just as in Santamato's) this is a result that follows from the equations of motion.

\subsection{The quantum action and its variation}

We now append the classical Lagrangian by a curvature term $\gamma(n)\lambda^2R[g,\phi]$, where $\lambda$ is a coupling constant that is ultimately fixed by observation. The term $\gamma(n)=\frac{1}{8}\frac{n-2}{n-1}$ is a numerical constant dependent on the dimensionality of the configuration space, chosen for later convenience.\footnote{Note that this expression for $\gamma$ is introduced with the knowledge of hindsight. Without this hindsight, we might have absorbed any such constant into $\lambda^2$.} That is, the action for the Trajectory-Weyl theory (TWT) is
\begin{equation}  \label{QMAction}
  I_{TWT}= \int\limits_\mathcal{C}d^nx\, \sqrt{g}\mu(x) \int\limits_{t_0}^{t_1} dt\,\Bigg( L_{class}(q,\dot{q},t) - \DERIV{S}{t}(q,t) + \gamma(n)\lambda^2R[g_{ij}(q,t),\phi_i(q,t)]\Bigg),
\end{equation}
where
\begin{equation}
 R[g_{ij}(q,t),\phi_i(q,t)] \equiv R_{Riem}[g_{ij}] + (n-1)\left[ (n-2)\phi_i\phi^i-\frac{2}{\sqrt{g}}\partial_i\left(\sqrt{g}\phi^i\right) \right] \label{WeylR}
\end{equation}
as obtained above. Note that we have included $\sqrt{g}$ in the volume element inside the integral since we allow for a general metric $g_{ij}$. The dependence of the variables on the initial position $x$ has been suppressed.

In what follows, we assume for simplicity that the configuration-space $\mathcal{C}$ is flat in the Riemannian sense ($R_{Riem}=0$) and the $\sqrt{g}$ is merely determined by coordinates. We take note though that in classical mechanics $\mathcal{C}$ may have a non-zero $R_{Riem}$ for systems with certain types of constraints which lead to a reduced configuration space (a $(n-l)$-dimensional surface with non-vanishing curvature within the $n$-dimensional configuration space, where $l$ is the number of independent constraints). Whether or not the `quantisation' method proposed here could (or should) be applied to a classical reduced configuration space in the presence of constraints is to be investigated in the future. 

The similarity of \ref{QMAction} with the Einstein-Hilbert action is obvious and may in part be the appeal of this approach. However, note once again that our action is defined in configuration space, not physical space(-time).

Let us for the sake of specificality consider the classical starting point to be an $N$-particle system in three dimensions ($n=3N$) with particle masses $m_I$, $I=1,\dots N$. The masses can be absorbed in the metric $g_{ij}$ on $\mathcal{C}$ in the usual way. For example for two particles of masses $m_1$ and $m_2$, we would take $g_{ij}=diag(m_1,m_1,m_1,m_2,m_2,m_2)$ (in Cartesian coordinates), enabling us to write terms in the form $\sum_I\half m_1\dot{\vec{q}}_I\cdot\dot{\vec{q}}_I = \half g_{ij}\dot{q}^i\dot{q}^j = \half \dot{q}^i\dot{q}_i$ and $\sum_i\frac{1}{2m_I} \vec{p}_I\cdot\vec{p}_I = \half g^{ij}p_ip_j=\half p^ip_j$, and so on, being careful to keep track of the natural index position for the dynamical quantities. See e.g.\ \cite{Lanczos1949} or any other book on analytic mechanics for further details. Hence there is no need to include explicit mass terms in what follows.

\paragraph{Variation of $q$: The law of motion}
Since $R$ has only dependence on $q$ and not $\dot{q}$, the boundary equation is unchanged, $p_i=\partial_iS$, and the Euler-Lagrange equation simply acquires an extra term $\gamma\lambda^2\PD{R}{q^i}$:
\begin{equation} 0 = \deriv{t}\PD{L_{class}}{\dot{q}^i}\big(q(t,x),\dot{q}(t,x),t\big)-\PD{L_{class}}{q^i}\big(q(t,x),\dot{q}(t,x),t\big)
			  -\gamma\lambda^2\PD{R}{q^i}\big[g_{ij}(q(t,x),t),\phi_i(q(t,x),t)\big] \label{QuantumEOM}
\end{equation}
The new term will ultimately be interpretable as a `quantum potential' $Q$ that matches Bohm's, as we will see. The equation then takes the form of the classical equation of motion with $Q$ appearing in the role of a new potential term, at least formally.

\paragraph{Variation of $\phi$: Relationship between $\phi$ and $\mu$}
We note that only the new term carries dependence on the Weyl one-form $\phi$. Hence we wish to minimise
\begin{align} I_R 
 &= (n+1)\int\limits_\mathcal{C} d^nx\sqrt{g}\mu\,\left[(n-2)\phi_i\phi^i-\frac{2}{\sqrt{g}}\partial_i(\sqrt{g}\phi^i)\right] \notag\\
 &= (n+1)\int\limits_\mathcal{C} d^nx\,\left[\mu\sqrt{g}(n-2)\phi_i\phi^i+\sqrt{g}\phi^i\cdot2\partial_i\mu\right].\label{OnlyRAction}
\end{align}
where we have partially integrated in order to arrive at the last line. The boundary term drops out since $\mu$ is normalisable and therefore vanishing at infintiy. For better readibility arguments have been suppressed. 

Following \citep{Santamato1984}, a little algebra reveals that the term to be extremised can be written as
\begin{align} I_R
 &= -\frac{n-1}{n-2}\int\limits_\mathcal{C}d^nq\,\mu\sqrt{g}g^{ij}\partial_i(\ln\mu)\partial_j(\ln\mu) \notag\\
 &\qquad	+\frac{n-1}{n-2}\int\limits_\mathcal{c}\mu\sqrt{g} g^{ij}\big((n-2)\phi_i+\partial_i(\ln\mu)\big)\big((n-2)\phi_j+\partial_j(\ln\mu)\big).
\end{align}
This brings $I_R$ into a form that is independent of derivatives of $\phi_i$. Furthermore, the first term contains no dependence on $\phi_i$ and can therefore be ignored for the purposes of extremisation.

The extremisation is easy since the integrand is positive definite (with $\sqrt{g}$ and $\mu$ being positive definite and the rest being of the square form $g^{ij}W_iW_j$ with $W_i=(n-2)\phi_i+\partial_i(\ln\mu)$). Hence the the minimum must be bounded from below by zero and indeed it is clear that this minimum is obtained by 
\begin{equation} \phi_i = -\frac{1}{n-2}\partial_i(\ln\mu). \label{phisolution}\end{equation}

This shows that the Weyl geometry is \emph{integrable}, that is, there is a scalar function $f$ on $\mathcal{C}$ such that $\phi_i=\partial_if$. Hence a local scale may be defined on $\mathcal{C}$ and the final length of a vector transported between two points in $\mathcal{C}$ is path-independent. It is worth noting that an integrable Weyl geometry allows a gauge choice such that $\phi_i=0$ (that is, $f=$ const.) everywhere.

\paragraph{The continuity equation}

The continuity equation may be obtained in the same manner as for the classical case, using the fact that the momentum is given by a spatial derivatives at all times as a result of the boundary equation \ref{eq:BoundaryEq.}. The presence of the curvature term has no impact. Allowing for arbitrary coordinates, it takes the form
\begin{equation} \label{CE} \pd{t}\left(\sqrt{g}\mu\right)+\partial_i\left(\sqrt{g}\mu\dot{q}^i\right) = 0. \end{equation}
The appearance of $\sqrt{g}$ with $\mu$ corresponds to the fact that $\mu$ is a scalar density of weight one. When rectilinear coordinates are used (where $g_{ij}$ is constant), then the factors of $\sqrt{g}$ may be dropped.

From the meaning of $\mu$, introduced in the process of taking the continuum limit for a configuration-space ensemble, we should have expected such a continuity equation to hold. It confirms that $\mu$ may be interpreted as a density. This equation furthermore shows that $\mu$ is conserved along a trajectory.

\subsection{Equivalence with equilibrium de~Broglie-Bohm theory}
Substituting the solution \ref{phisolution} into the expression \ref{WeylR} for the curvature, we find that (by simple algebra, and using $\gamma=\frac{1}{8}\frac{n-2}{n-1}$)
\begin{align} R
 &= R_{Riem} +\frac{n-1}{n-2}\left[\partial_i(\ln\mu)g^{ij}\partial_j(\ln\mu)+\frac{2}{\sqrt{g}}\partial_i\left(\sqrt{g}g^{ij}\partial_j(\ln\mu)\right)\right]
			      \notag \\
 &= R_{Riem} +\frac{1}{2\gamma\sqrt{\mu}}\partial_i\left(\sqrt{g}g^{ij}\partial_j\sqrt{\mu}\right).
\end{align}

With this, the Hamilton-Jacobi equation derived from the Lagrangian $L-\DERIV{S}{t}+\gamma\lambda^2R$ becomes
\begin{align} 0 
  &= \PD{S}{t} + H(q,\nabla S,t) \notag\\
  &= \PD{S}{t} + H_{class}(q,\nabla S,t)- \gamma\lambda^2R_{Riem}  -\frac{\lambda^2}{2}\frac{\partial_i(\sqrt{g}g^{ij}\partial_j\sqrt{\mu})}{\sqrt{\mu}},
	\label{QHJE}
\end{align}
where we used the fact that the momentum conjugate to $\phi_i$ vanishes identically since $\dot{\phi}_i$ does not appear in the expression for $R$.

The reader familiar with de~Broglie-Bohm theory in Bohm's second-order formulation will recognise the last term on the right as the quantum potential (at least for a broad class of particle systems), 
provided $\lambda^2 = \hbar^2$ (which ultimately is just the experimental discovery of the value of a constant, or, viewed differently, the discovery of a property of the system of units used) and that $\mu$, the configuration-space measure, can be interpreted as the equilibrium density $|\Psi|^2$. But the measure $\mu$ was introduced as describing the local (in $\mathcal{C}$) density of systems of our ensemble in the continuum limit and so this is indeed the natural interpretation. 

In order to obtain the evolution of the ensemble as a whole, eq.\ \ref{QHJE} must now be solved simultaneously with the contiuity equation (\ref{CE}) for $\mu$ and $S$, which in practice would be done by combining them into the complex field $\psi=\sqrt\mu e^{iS/\hbar}$ on $\mathcal{C}$. Solving for a single trajectory alone is not possible, since $R$ contains reference to $\mu$, which is determined by the collection of all trajectories.

The equations \ref{QHJE} and \ref{CE} match, of course, the usual quantum Hamilton-Jacobi and continuity equations known from equilibrium de~Broglie-Bohm theory, with the difference that no reference to any wavefunction $\psi$ is made and the `quantum potential' contains $\sqrt{\mu}$ instead of $|\psi|$, with the consequence that there is no state of quantum non-equilibrium.\footnote{The term 'non-equilibrium' here is meant in the sense of de~Broglie-Bohm, that is, a state where $\rho\neq|\psi|^2$. This is impossible here since $\psi$ is not an ontological entity in its own right but only definable in terms of $\rho$ and $S$ (up to an unphysical overall phase factor). Contrast this to `many-de~Broglie-Bohm worlds' theory \citep{Sebens2014,Valentini2010}, where such $\rho$ and $|\psi|$ are a priori independent of one another.}

Note that de~Broglie's first-order dynamics also emerges from this formalism (at least for certain Lagrangians), namely as the condition $p_i=\partial_iS$ derived from the boundary equation resulting from the open-endpoint variation.

Thus the equations derived here are equivalent with those of de~Broglie-Bohm theory. It appears that therefore the predictions of quantum mechanics have also been recovered. However, this conclusion would be incorrect, as we will discuss in the next section.

\section{Nodes and the phase-like nature of \texorpdfstring{$S$}{S}}\label{Wallstrom}
In this section we discuss the inequivalence between the present theory as it stands and de~Broglie-Bohm quantum mechanics. The issue is related to the respective single versus multi-valuedness of $S$. However, we show how the phase-like behaviour of $S$ can be recovered in our theory using a further generalisation of our Lagrangian. To do so, we first discuss the concept and nature of `nodes' in our theory and their relationship to the geometry. This requires a shift in our perspective: from a configuration-space manifold \emph{with} trajectories to a manifold \emph{of} trajectories.

\subsection{Wallstrom's criticism}
The problem with the above theory, and indeed any many-trajectories theory with or without a geometric foundation \citep{Sebens2014,HallDeckertWiseman2014,SchiffPoirier2011} is that the space of solutions of such a theory is smaller than that of de~Broglie-Bohm quantum mechanics.\footnote{Alternatively a many-trajectories theory may be constructed using the velocities or momenta themselves as the variables and not making reference to a function $S$. In this case the space of solutions is not smaller than that of de~Broglie-Bohm but instead larger since angular momenta can take continuous values. Hence there are two versions of Wallstrom's criticism applicable to different sorts of formulations.} While on a small local patch in configuration space the equations have the same solutions, a difference in the possible \emph{global} properties of the solutions arises. To our knowledge this was first noted by Wallstrom \citep{Wallstrom1994} in the context of the Madelung hydrodynamic approach.

The function $S$, when introduced as the phase of the (physically real) wavefunction $\psi=|\psi|e^{iS/\hbar}$, is multi-valued, $S\sim S+m\cdot2\pi\hbar$, $m\in\mathbb{Z}$. Furthermore, $S$ is \emph{undefined} at nodes of the wavefunction (points where $\psi=0$). These two facts together imply that there may be `states' (functions $\psi$) such that a closed loop $L$ may be found such that
\begin{equation} \oint_L \partial_iS\,dl^i \neq 0, \label{eq:non-integrability}\end{equation}
where $dl$ is a line element in the configuration space. Specifically this is possible if and only if $L$ is \emph{non-contractible}, in the sense that it is impossible to deform $L$ continuously into a point without passing through a node. Such loops exist in the presence of $(n-2)$-dimensional nodes (points in two dimension, lines in three dimensions, etc.). 

Contrast this to the case where $S$ is introduced as a scalar function on $\mathcal{C}$ \citep[e.g.]{SantamatoMartini2014}. Here $S$ is defined even where $\mu=0$ and is a real field on $\mathcal{C}$, so that the integral \ref{eq:non-integrability} vanishes for \emph{any} loop.

Solutions to the Schr\"odinger equation that display this non-integrability of $\partial_iS$ arise, for example, in the case of stationary states of hydrogen-like atoms with non-zero angular momentum. It follows that the present theory and other, similar many-trajectories theories do not allow for non-zero angular momentum states and therefore do not match the predictions of quantum mechanics.

Wallstrom's criticism has remained largely unaddressed by authors proposing many-trajectory theories, as we discussed in the short survey above. 
The solution we will construct here is superficially similar to Smolin's \citep{Smolin2015}, who introduces a unimodular complex variable $w$ (a `complex factor beable') in place of the momentum. However, in our case the goal will be achieved via a generalisation of the total derivative appearing in the Lagrangian, not by revising the mathematical representation of momenta.

\subsection{Nodes}
The necesssity of having nodes present in cases where this mismatch between our `Trajectory-Weyl theory' (TWT) and quantum mechanics arises leads us to investigate the nature of nodes in the former. The term `node' in TWT is, of course, not a good choice of terminology given that no waves are involved. However, it is clear what is meant: A point $P$ in $\mathcal{C}$ is a node if $\mu(P)=0$. 
In de~Broglie-Bohm theory this does not necessarily imply that there is no trajectory that passes through the node. Whether a consistent law of motion exists for this system depends on whether the phase function $S(q)$ or at least the velocity field can be analytically extended through the node. This is trivial if the wave function is real (e.g. the $n=2$, $\ell=1$, $m=0$ state of hydrogen), in which case the particle placed at the node remains at the node. On the other hand, the $n=2$, $\ell=1$, $m=1$ state does not allow for such an analytic extension. There can be no continuous velocity field that includes the node and no trajectory can exist there within the framework of the theory. More generally, a trajectory exists (in a mathematical sense) at the node if and only if the loop integrals $\oint_L \partial_iS\,dl^i$ around the node vanish. Whether or not this extra `world' exists in an ontological sense is however irrelevant for the dynamics of all other `worlds' in the case of `many-de~Broglie-Bohm-worlds' theory. Furthermore, adding a single trajectory does not change the local system density.

Above we obtained the result $\phi_i=-\frac{1}{n-2}\partial_i(\ln\mu)$ (eq.\ \ref{phisolution}). At a node, where $\mu=0$, this becomes undefined. That is, the mathematical notion of vector transport and the idea of `scale' fails to make sense at nodes. So nodes correspond to a certain kind of geometric singularity.

Let us examine the behaviour of the Weyl form $\phi_i$ in the immediate vicinity of a node. Suppose $\mu\sim r^{2m}$ as we approach the node, where $r$ is the distance from the node and $m$ is some (not necessarily integer) positive power. The factor of two in the exponent is for later convenience.

Approaching the node radially, that is, moving in the negative-$r$ direction, we see that
\begin{equation} \phi_{-r}=-\frac{2}{n-2}\pd{(-r)}\ln r^{2m} = \frac{2m}{n-2}\frac{1}{r}, \end{equation}
which shows that the length of a vector being transported radially towards the node by a small distance $\delta r$ grows by a factor proportional to $r^{-1}$. Since the Weyl geometry is integrable, we can also look at the length of a vector itself, which goes as
\begin{equation} \ell\sim -\ln r, \end{equation}
and so $\ell\rightarrow\infty$ as $r\rightarrow0$.

\subsection{From a manifold with trajectories to a manifold of trajectories}

It is now time for a shift in our thinking. We began by considering systems moving in the configuration-space manifold $\mathcal{C}$ with time, forming trajectories in the `configuration-space-time' ($\mathcal{C}\,\otimes$ time, sometimes called the `extended configuration space' \citep{Lanczos1949}). However, the dynamics of the theory imply that geometric properties are undefined at certain points in $\mathcal{C}$, namely those points not occupied by a system. This suggests that instead of considering systems particles (or `corpuscles', or `worlds') moving \emph{in} a manifold, we should consider the manifold \emph{of} all system particles (or of all trajectories if we think of it in a configuration-space-time picture), which we shall call $\mathcal{T}$. The configuration space merely becomes a way to label systems within our continuously infinite ensemble, namely by using the physical properties (the value of the variables $q_i$). In the presence of nodes, $\mathcal{T}$ is not simply connected.

The view taken is therefore the following. The physical world in its entirety consists of an infinite collection of similar systems (described by the same set of variables, that is, corresponding to points in the same configuration space) whose properties are specified in the form of $n$ variables $q^i$. These variables determine the continuity properties of the collection, thereby forming an $n$-dimensional manifold $\mathcal{T}$. Since the number density of systems is not uniform across the $q^i$, this manifold, in general, has non-zero curvature. The manifold $\mathcal{T}$ may be embedded in a configuration space $\mathcal{C}$ of the same dimension via a Weyl geometry, on which points are labelled by the $q^i$. Continuous subsets of $\mathcal{C}$ with a dimension smaller than $n$ might not have an inverse image (are not mapped to in the embedding) and form nodes, illustrating that the original manifold might not be simply connected. In practice, we work in $\mathcal{C}$, labelling systems by $q^i$ and tracking their local number density via the function $\mu:\mathcal{C}\rightarrow\mathbb{R}_{\geq0}$.

The function $S$ introduced in the action as an arbitrary function is defined on $\mathcal{T}$, not $\mathcal{C}$. Following the embedding in $\mathcal{C}$ it is therefore natural that $S$ is not defined at nodes, since these are exactly those points that are \emph{not} in the image of $\mathcal{T}$. Of course, by itself this does not answer Wallstrom's criticism. While the presence of nodes is necessary for a mismatch between na\"ive TWT and quantum mechanics to occur, it is not sufficient. In quantum mechanics $S$ is multi-valued and the presence of nodes merely allows one to find closed paths changing $S$ continuously from one value at a point to another at the same point.

Before coming to the second part of our solution to Wallstrom's problem, return briefly to the question of the existence of trajectories at nodes. In TWT, unlike in `many-de~Broglie-Bohm worlds' theory, the manifold of system particles is considered fundamental. This manifold may or may not have a non-trivial topology, and what functions $S$ (or `$U+W$' as introduced in the next section) are possible is contingent on that topology. One result of this is that a scenario in which there is a node in the wave function (as described by the standard theory) through which $S$ can be analytically extended may or may not correspond to a simply connected topology for $\mathcal{T}$. But whether or not there is a system particle at the node matters ---unlike in the case of `many-de~Broglie-Bohm worlds' --- since it changes the topology of the manifold and therefore the global properties of functions definable on it.

\subsection{The most general time derivative \texorpdfstring{$dS/dt$}{dSdt}}

We now discuss a possible solution to Wallstrom's problem. It is based on the notion that is desirable to have the most general Lagrangian that gives the correct equations of motion. 
We therefore wish to add the most general total time-derivative $dS/dt$. This is, in fact, the sum of two time derivatives, one of which takes values on $\mathbb{R}$ as usual, the other one is a derivative of a function that takes values from the \emph{closed} one-dimensional manifold $S^1$, not $\mathbb{R}$.  

In order to see this, let us take a step back and consider how $S$ was introduced. We added an arbitrary total derivative $\DERIV{S}{t}$ to the action. The only requirement was that $\DERIV{S}{t}\in\mathbb{R}$. This requirement may seem almost not worth stating since it is generally assumed that terms in the action take values over $\mathbb{R}$ or sometimes $\mathbb{C}$. However, the implication of $\DERIV{S}{t}\in\mathbb{R}$ is \emph{not} that $S$ itself must take values on $\mathbb{R}$. Instead we could define $S:\mathcal{T}\rightarrow S^1$, that is, let $S$ map from $\mathcal{T}$ (or equivalently $\mathcal{C}\setminus\{$nodes$\}$) to the $\emph{closed}$ one-dimensional manifold, the `circle' $S^1$. The spaces $S^1$ and $\mathbb{R}$ share the same tangent space $\mathbb{R}$. Furthermore, they are the \emph{only} spaces that do so and our Lagrangian will therefore indeed be the most general.

It turns out that the resultant function `$S$' has the required propeties to resolve Wallstrom's problem. 
A critic might be inclined to say that we `cheated' by giving $S$ the required phase-like property by fiat. 
Yet without wishing to discuss the philosophical issues too deeply, this is no different than any other introduction of mathematical objects in quantitative physics. The choice of what particular type of object should represent some proposed physical entity is founded on the properties that entity is supposed to have, inferred from observed phenomena and, arguably, certain theoretical guiding principles (such as generality, simplicity, symmetries, etc.). For example, in standard quantum mechanics we invent a mathematical object called the `wave function' that takes values from $\mathbb{C}$ on configuration space (or, more abstractly, some Hilbert space whose physical meaning is more obscure). 
Similarly, we have chosen the correct mathematical structure to address Wallstrom's objection and match the predictions of the theory to empirical facts.

We could justify the step of adding to the action a total time-derivative of a function $S:\mathcal{T}\rightarrow S^1$ rather than one mapping to $\mathbb{R}$ by its ultimate observational success (such as in measurements of atomic angular momentum). However, since the requirement $\DERIV{S}{t}\in\mathbb{R}$ gives two options, $S:\mathcal{T}\rightarrow\mathbb{R}$ and $S:\mathcal{T}\rightarrow S^1$, and neither changes the equations of motion, a priori we should add \emph{both} terms in order to achieve maximal generality. The added term is therefore not an \emph{ad-hoc} addition, but a natural generalisation. Let us therefore introduce a function $U:\mathcal{T}\rightarrow\mathbb{R}$, while the symbol $W$ will be used for the function $W:\mathcal{T}\rightarrow S^1$. These take the place of the symbol '$S$'. Our action becomes
\begin{equation}  \label{QMActionWithDoubleTDs}
  I_{TWT}^\star=\int_{t_0}^{t_1} dt\, \int_\mathcal{C}d^nx\, \sqrt{g}\mu(x) \Bigg( L_{class}(q,\dot{q},t) + \gamma(n)\lambda^2R[g_{ij}(q,t),\phi_i(q,t)] - \DERIV{U}{t}(q,t) - \DERIV{W}{t}(q,t) \Bigg).
\end{equation}
The equation of motion (\ref{QuantumEOM}) remains unchanged, as does the result for the Weyl geometry (\ref{phisolution}). The one difference is in the boundary equation, which now reads
\begin{equation} p_i = \partial_i W + \partial_i U. \end{equation}
All the previous derivations to recover de~Broglie-Bohm trajectories hold provided we replace $S$ by `$W+U$'. The expression is in quotation marks since addition is not defined for two functions taking values in different spaces, even if both are one-dimensional, and is therefore not to be interpreted as literal addition. However, $S=`U+W'$ never occurs itself, only its derivatives do and among them addition is well-defined (since they all take values in $\mathbb{R}$).

Now since $U$ is a globally defined real field, we have 
\begin{equation}\oint_L\,\partial_i U\cdot dl^i =0\end{equation}
for any closed loop, so there is no contribution from $U$ to closed-loop integrals of $\nabla S$. Meanwhile,
\begin{equation}\oint_L\,\partial_i W\cdot dl^i = z\cdot h,\qquad z\in\mathbb{Z},\end{equation}
where $h$ is defined to be the value obtained when the image of the loop $L$ under $W$ winds around the manifold $S^1$ exactly once. The numerical value of $h$ depends on how the manifold $S^1$ is parametrised (how it is represented in terms of real numbers, ultimately a choice of units). The value of $h$ is the same for all nodes in the manifold.

The constant $h$ can be thought of as the `circumference' of $S^1$ if we visualise the manifold as a circle embedded in $\mathbb{R}^2$. It functions as a `constant of conversion' between lengths on $S^1$ and on $\mathbb{R}$. For dimensional reasons it takes units of angular momentum if dimensionful variables are used. What is left to be addressed is the relationship between the associated radius, $\hbar=h/2\pi$, and the geometric `coupling constant' $\lambda$.

The proportionality of the geometric coupling constant $\gamma\lambda^2$ follows from dimensional analysis. The functions $U$ and $W$ must have the same dimension as the action, length squared over time ($L^2T^{-1}$). The only fundamental scale in the theory is the `radius' $\hbar$, which therefore must also have dimensions $L^2T^{-1}$. But the dimensions of the curvature $R$ are $L^{-2}$, so $\gamma\lambda^2$ must have the dimensions of $\hbar^2$, allowing us to set $\lambda=\hbar$, leaving only the numerical constant $\gamma$. 

The constant $\hbar$ is the one fundamental scale provided by the theory. It is remarkable that this one fundamental scale is, in fact, sufficient to compare all terms in the Lagrangian (that is, to let them have the same dimensionality). It is even possible to go one step further and demand that the action be dimensionless. This can be achieved by rescaling $I_{TWT}$ by a factor $\hbar^{-1}$. In this case $U$ and $W$ are dimensionless, too, and the closed manifold onto which $W$ maps is simply parametrised by angular values.

So only the numerical value of $\gamma$ is left to be discussed. The value $\frac{1}{8}\frac{n-2}{n-1}$ was, of course, chosen with the hindsight of experimental outcomes (matching those of quantum mechanics). However, it is itself distinguished: Its $n$-dependence is the unique choice such that the resultant equations of motion do not explicitly depend on $n$. Only the geometric properties of the manifold do (eq.\ \ref{phisolution}). The coupling constant resemles a `conformal coupling` as familiar from field theory in Riemannian geometries.

\section{Conclusion}

In this paper we showed how the phenomenology of non-relativistic quantum mechanics may be recovered starting with an `ensemble' of systems continuously distributed in configuration space. Thus our theory is one of many worlds. By including a curvature term in the action the trajectories turned out to match those of equilibrium de~Broglie-Bohm theory. Our construction was closely modelled along that of \citep{Santamato1984}, although our starting point was different. Nor is the idea of recovering quantum dynamics via an infinite ensemble of systems, covering all possible configurations, completely new, although it is a relatively recent development.

Our purpose here was to show that the two ideas, a dynamical Weyl geometry together with a `many-trajectory' theory, form a natural union. In order to emphasise this point we showed how Wallstrom's objection, a key problem with `quantum' theories not based on a wavefunction, can be overcome via a generalisation of the total time derivative in the Lagrangian. 

In our procedure we demoted the configuration space to a purely mathematical concept, with the physical reality corresponding to a manifold of configuration-space particles (or `worlds'), or ---in a configuration-space-time picture--- of trajectories. Other than the fact that this manifold, unlike the configuration space, has topological properties that allow the existence of non-contractible loops and thereby forms the first step in our solution to Wallstrom's objection, it may also be philosophically preferable: In standard quantum mechanics the wavefunction is defined on configuration space and it is unclear what this space is without appeal to some previous classical theory with a certain set of dynamical variables. In contrast, the present Trajectory-Weyl theory has no such ambiguous ontology. Functions such as $S$ (or $U$ and $W$) are defined on the set of all worlds, each of which is physically concrete. If the manifold allows for non-contractible loops then the $S^1$-valued function $W$ may yield non-zero, quantised values for the integral $\oint \partial_iW\cdot dl^i$. The appearance of $W$ in the generalisation of the Lagrangian thus completes the solution to Wallstrom's objection.

The geometric approach has been extended to relativistic particle quantum mechanics \citep{Santamato1984b,Santamato1985} (in the sense of a non-field-theoretic picture, where the Klein-Gordon equation is considered to hold for a wavefunction, not for a classical field) and the particle Dirac equation \citep{SantamatoMartini2011,SantamatoMartini2014}, starting from the configuration space of a classical rigid body (although without attention to Wallstrom's criticism). However, whether or not such a geometric approach is ultimately viable depends on its success of recovering results of quantum field theory. 


Wallstrom's problem exists here, too. Recall that the presence of non-contractible loops requires the existence of nodes of codimension two. In field theory this occurs in states corresponding to particular linear combinations of two field modes, analogous to circular polarisation. For example, if only two such modes $k_1,k_2$ in Fourier space are non-zero, then the infinite-dimensional wave functional can be written in the form of a two-dimensional wave function. For example,
\begin{equation}\Psi(k_1,k_2) = R(|k_1^2+k_2^2|)\;[\cos k_1 + i\;\sin k_2],\end{equation}
where $R(|k_1^2+k_2^2|)$ is the `radial' part of $\Psi$ and satisfies $R(0)=0$, so that there is a node on the `$\infty-2$' dimensional `line' $k_1=0=k_2$. The existence of such states implies that Wallstrom's objection also applies to the infinite-dimensional field theory, not just to simple non-relativistic particle mechanics.

Whether the program of a geometric foundation of quantum field theory in this manner ultimately succeeds is unclear, although one may be hopeful. It will also be interesting to investigate if and how the approach may be extended to gravitational dynamics, possibly even a full theory of quantum gravity. In particular, if the techniques employed in this paper are used as a method of `quantising' the 3-metric field by considering an ensemble of trajectories in superspace, does one obtain a theory equivalent to `frozen' dynamics of the Wheeler-de~Witt equation, or does one find differences? This will be the subject of future investigation.

\section*{Acknowledgements}
We wish to thank Antony Valentini for bringing the Wallstrom problem to our attention, for various discussions of the subject and for a critical review of this manuscript, as well as Lucien Hardy for multiple helpful discussions during the early development of these ideas.


\bibliographystyle{ieeetr}	
\bibliography{../../Bibloi}	



\end{document}